# Identification of a complete YPT1 Rab GTPase sequence from the fungal pathogen *Colletotrichum incanum*

Cecil Barnett-Neefs and Ruth N. Collins

Dept. Molecular Medicine, College of Veterinary Medicine, Cornell University, Ithaca NY 14853

## Introduction

Colletotrichum represent a genus of fungal species primarily known as plant pathogens with severe economic impacts in temperate, subtropical and tropical climates (Cannon et al., 2012). Colletotrichum pathogens are useful as models for studying the molecular basis of the differentiated cell types associated with specialized structures associated with infection and fungal-plant interactions (Perfect et al., 1999). Consensus taxonomy and classification systems for *Colletotrichum* species have been undergoing revision as high resolution genomic data becomes available (O'Connell et al., 2012). Here we propose an alternative annotation that provides a complete sequence for a Colletotrichum YPT1 gene homolog using the whole genome shotgun sequence of *Colletotrichum incanum* isolated from soybean crops in Illinois, USA.

## Results and Discussion

A BLAST (Altschul et al., 1990) search for Rab GTPase YPT1-related protein sequences identified the hypothetical ORF OHW89796.1 in *Colletotrichum incanum*. Pairwise global alignment reveals the hypothetical ORF has 71.8% identity, 80.9% similarity with Ypt1p (Figure 1) however the database annotated ORF lacks a start methionine. The *Colletotrichum incanum* genome features consistent donor splice sites of "GT" and acceptor splice sites of "AG" in line with the majority of fungal introns (Kupfer et al., 2004) so we manually checked the genomic sequence for possible alternative splice sites. The annotated ORF is located within the genomic sequence of KV842198 and is encoded by 4 exons (Figure 2A). A hypothetical acceptor splice site is located adjacent to the first annotated exon of OHW89796, with a hypothetical donor splice site yielding a 191 bp intron (Fig. 2B). If these consensus splice sites constituted an authentic intron, the resulting mRNA would provide a $NH_2$-terminal methionine for translation (Figure 3A) and, moreover, the additional residues of the translated protein product (MNPE) would include a strong homology to the $NH_2$-terminus of Ypt1p in line with other fungal orthologs of this protein (Figure 3B).

In order to provide experimental evidence for an alternative OHW89796 exon, we searched the dBEST database. In order to determine if expressed *Colletotrichum* sequences would include the sequence homologous to the proposed additional $NH_2$-terminal peptide with start methionine we used a search query of the complete open reading frame (609 nucleotides, see Appendix) of the proposed OHW89796 ORF. The highest scoring sequence identified in this survey was FN670696, a cDNA clone representing mRNA from *Colletotrichum higginsianum* with 581/609 (95%) identity and zero gaps (Figure 4). Importantly, the expressed and experimentally identified sequence contains a complete reading frame including the proposed $NH_2$-terminal peptide and start methionine. These data support the suggestion that the annotation for genomic OHW89796 should be updated to include an additional exon. Accurate identification of genes and proteins relevant to fungal infection and pathogenesis will contribute to our understanding of the biology of *Colletotrichum* and solving the problem of infection control.

## Methods

Pairwise global protein alignments were performed using the EMBOSS (Rice et al., 2000) needle 6.6.0 webserver with EBLOSUM62 matrix, gap penalty 10.0, extension penalty 0.5. DNA sequence alignments used default parameters.

SnapGene Viewer Version 4.1.9 was used for the graphical representation of DNA maps.

OHW89796 was analysed with nucleotide BLAST using the Expressed Sequence Tags database (https://www.ncbi.nlm.nih.gov/nucest) and default parameters.

Figure Legends

Figure 1

Global protein alignment of the Rab GTPase Ypt1p from Saccharomyces cerevisiae with the database annotated ORF from OHW89796.1 in Colletotrichum incanum. The annotated ORF is lacking a start methionine.

Figure 2

A. Database annotated ORF OHW89796 from KV842198 contains 4 exons. KV842198 is 22,107 bp for clarity only the relevant portion of the genomic sequence is depicted. Numbering is according to the KV842198 database sequence.

B. Proposed identification of upstream exon for complete *Colletotrichum* YPT1 homolog sequence.

Figure 3

A. DNA sequence map surrounding location of proposed intron and translational start site. Numbering is from KV842198.

B. Exon 1 encoded $NH_2$-terminal peptide (outlined in black) compared to Ypt1p. This peptide contains a level of homology to Ypt1p consistent with the homology observed throughout the ORF (Fig. 1)

Figure 4.

*Colletotrichum higginsianum* FN670696 IMI349063A biotrophic infection *Colletotrichum* cDNA sequence from the EST database showing alignment with the proposed complete DNA sequence encoding the ORF for the *Colletotrichum incanum* YPT1 ortholog.

# Appendix

Complete nucleotide and protein sequences proposed for *Colletotrichum incanum* OHW89796.

Figure 1

```
YPT1         1 MNSEYDYLFKLLLIGNSGVGKSCLLLRFSDDTYTNDYISTIGVDFKIKTV  50
               ||||||||||||:|||||||||||||:|||||..||||||||||||:|:
OHW89796.1   1 ----YDYLFKLLLIGDSGVGKSCLLLRFADDTYTESYISTIGVDFKIRTI  46

YPT1        51 ELDGKTVKLQIWDTAGQERFRTITSSYYRGSHGIIIVYDVTDQESFNGVK 100
               |||||||||||||||||||||||||||||||:|||.:||||||.:|||.||
OHW89796.1  47 ELDGKTVKLQIWDTAGQERFRTITSSYYRGAHGICVVYDVTDMDSFNNVK  96

YPT1       101 MWLQEIDRYATSTVLKLLVGNKCDLKDKRVVEYDVAKEFADANKMPFLET 150
               .||||||||||..|.|||||||.|:.||:||||.||||||||:...:|||||
OHW89796.1  97 QWLQEIDRYATEGVNKLLVGNKSDMSDKKVVEYTVAKEFADSLGIPFLET 146

YPT1       151 SALDSTNVEDAFLTMARQIKESMSQQNLNETTQKKEDKGNVNL-KGQSL- 198
               ||.:::|||.||||||||||||.|.....|.|        |.:|.: .|..:
OHW89796.1 147 SAKNASNVEQAFLTMARQIKERMGSTTANNT------KPSVQVGPGHGVS 190

YPT1       199 TNTGGGCC*     207
               :|:|||||
OHW89796.1 191 SNSGGGCC*     199
```

Figure 2

A. database annotated ORF OHW89796 from KV842198

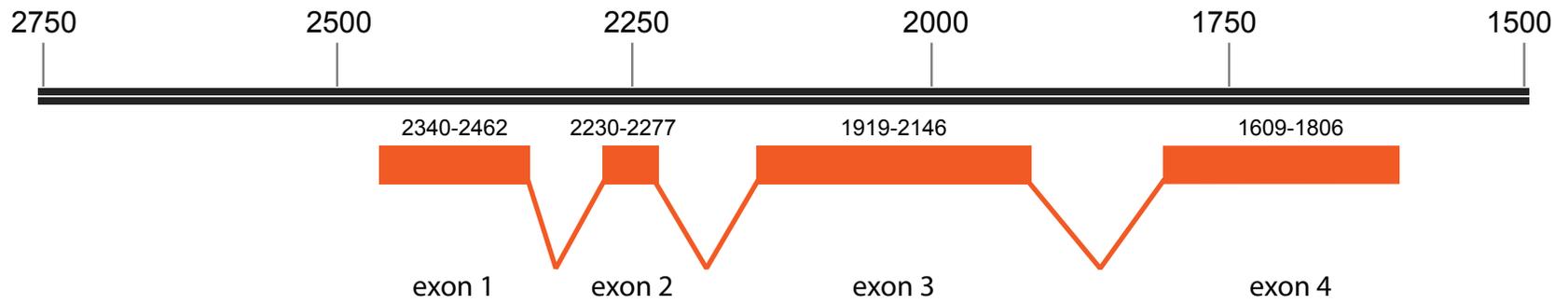

B. proposed alternative ORF OHW89796 from KV842198

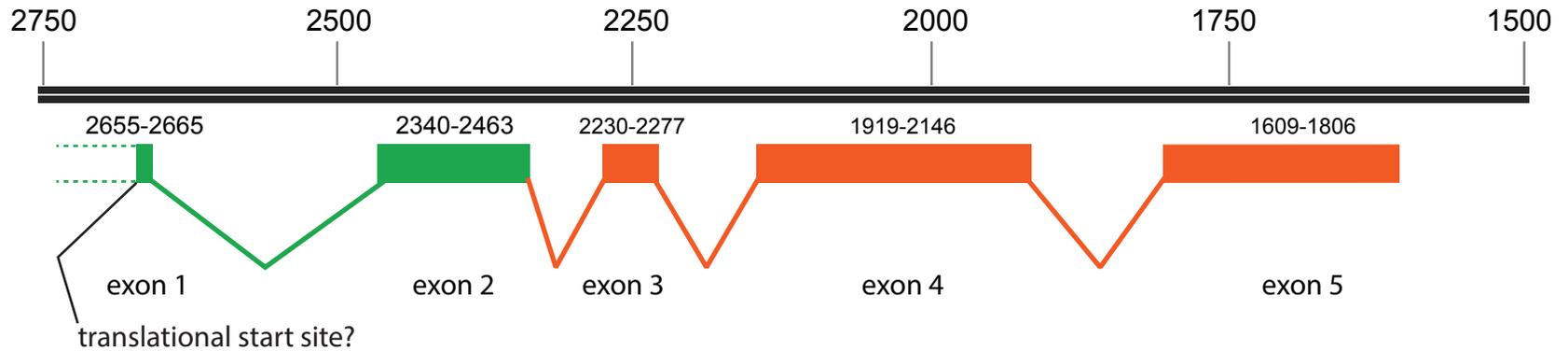

Figure 3

A.

```
KV842198    2670                              2460            2450
  5' atgaaccccga gt              ag gtacgactacctt
     |  |  |  |  |  |  |  |  |  |  |- - - - - - -|  |  |  |  |  |  |  |  |  |  |
  3' tacttggggct ca              tc catgctgatggaa
     proposed           intron   intron
     translational
     start site
```

⬇

```
  5' atgaaccccgagtacgactacctt
     |  |  |  |  |  |  |  |  |  |  |  |  |  |  |  |  |  |  |  |  |  |  |
  3' tacttggggctcatgctgatggaa
      M   N   P   E   Y   D   Y   L
```

B.

```
YPT1              1 MNSE
                    ||:|
OHW89796.1        1 MNPE
```

Figure 4. Alignment of proposed *Colletotrichum incanum* YPT1 ortholog OHW89796 with cDNA clone from *Colletotrichum higginsianum* FN670696

```
OHW89796    1   ATGAACCCCGAGTACGACTACCTTTTCAAGCTCCTCCTCATCGGTGACTCCGGTGTTGGA   60
                ||||||||||||||||||||||||| ||||||||| ||||||||||||||||||||||||
FN670696   95   ATGAACCCCGAGTACGACTACCTCTTCAAGCTTCTCCTCATCGGTGACTCCGGTGTTGGA  154

OHW89796   61   AAGTCTTGCCTTCTGTTGCGTTTCGCCGACGACACCTACACTGAGTCCTACATCTCCACC  120
                ||||| |||||||||||||||| |||||||||||||||||||||||||||||||||||||
FN670696  155   AAGTCCTGCCTTCTGTTGCGATTCGCCGACGACACCTACACTGAGTCCTACATCTCCACC  214

OHW89796  121   ATCGGTGTCGACTTCAAAATCCGCACGATAGAGCTCGATGGAAAGACGGTGAAACTTCAG  180
                || ||||||||| |||||||||||||||||||||||||||||||||||||||||| |||
FN670696  215   ATTGGTGTCGATTTCAAAATCCGCACGATAGAGCTCGATGGAAAGACGGTGAAGCTGCAG  274

OHW89796  181   ATCTGGGATACTGCCGGCCAGGAGCGTTTCCGCACCATCACTTCCTCTTACTACCGCGGT  240
                |||||||| || || |||||||||||||||||||||||| | ||||||||||||| |||
FN670696  275   ATCTGGGACACCGCTGGCCAGGAGCGTTTCCGCACCATCGCCTCCTCTTACTACCGTGGT  334

OHW89796  241   GCTCACGGCATCTGCGTCGTCTACGATGTGACCGACATGGACTCGTTCAACAACGTCAAG  300
                || |||||||||||||||||||||||||||||||||||||||||| |||||||||| |||
FN670696  335   GCCCACGGCATCTGCGTCGTCTACGATGTGACCGACATGGACTCCTTCAACAACGTTAAG  394

OHW89796  301   CAGTGGCTCCAGGAGATTGACAGATATGCCACCGAGGGCGTCAACAAGCTGCTTGTCGGC  360
                |||||||||||||||||||||||| |||||||||||||||||||||||||||| ||||||
FN670696  395   CAGTGGCTCCAGGAGATTGACAGATACGCCACCGAGGGCGTCAACAAGCTGCTCGTCGGC  454

OHW89796  361   AACAAGAGCGATATGTCCGACAAGAAGGTTGTCGAGTACACCGTCGCCAAGGAGTTCGCC  420
                |||||||||||||||||||||||||||||||||||||||||||||||||||||||| |||
FN670696  455   AACAAGAGCGATATGTCCGACAAGAAGGTTGTCGAGTACACCGTCGCCAAGGAGTTTGCC  514

OHW89796  421   GACAGCCTGGGCATCCCTTTCCTCGAGACCTCTGCGAAGAACGCCAGCAACGTCGAGCAG  480
                |||||||||||||||||| |||||||||||||||||||||||||||||||||| ||||||
FN670696  515   GACAGCCTGGGCATCCCCTTCCTCGAGACCTCTGCGAAGAACGCCAGCAACGTTGAGCAG  574

OHW89796  481   GCTTTCCTGACCATGGCCCGCCAAATCAAGGAGCGCATGGGCAGCACGACCGCGAACAAC  540
                |||||| |||||||||||||||||||||||||||||||||||||||||||||| ||||||
FN670696  575   GCTTTCTTGACCATGGCCCGCCAAATCAAGGAGCGCATGGGCAGCACGACCGCCAACAAC  634

OHW89796  541   ACGAAACCCTCTGTGCAGGTTGGGCCCGGACACGGCGTCTCCTCGAACTCGGGCGGCGGC  600
                |||||||||| |||||||| || |||||||||||||||||| ||||||||||||||||||
FN670696  635   ACGAAACCCTCCGTGCAGGTCGGACCCGGACACGGCGTCTCTTCGAACTCGGGCGGCGGC  694

OHW89796  601   TGCTGCTAA   609
                |||||||||
FN670696  695   TGCTGCTAA   703
```

```
atgaaccccgagtacgactaccttttcaagctcctcctcatcggtgactccggtgttggaaagtcttgccttctg   75
 M  N  P  E  Y  D  Y  L  F  K  L  L  L  I  G  D  S  G  V  G  K  S  C  L  L

ttgcgtttcgccgacgacacctacactgagtcctacatctccaccatcggtgtcgacttcaaaatccgcacgata  150
 L  R  F  A  D  D  T  Y  T  E  S  Y  I  S  T  I  G  V  D  F  K  I  R  T  I

gagctcgatggaaagacggtgaaacttcagatctggatactgccggccaggagcgtttccgcaccatcacttcc  225
 E  L  D  G  K  T  V  K  L  Q  I  W  D  T  A  G  Q  E  R  F  R  T  I  T  S

tcttactaccgcggtgctcacggcatctgcgtcgtctacgatgtgaccgacatggactcgttcaacaacgtcaag  300
 S  Y  Y  R  G  A  H  G  I  C  V  V  Y  D  V  T  D  M  D  S  F  N  N  V  K

cagtggctccaggagattgacagatatgccaccgagggcgtcaacaagctgcttgtcggcaacaagagcgatatg  375
 Q  W  L  Q  E  I  D  R  Y  A  T  E  G  V  N  K  L  L  V  G  N  K  S  D  M

tccgacaagaaggttgtcgagtacaccgtcgccaaggagttcgccgacagcctgggcatccctttcctcgagacc  450
 S  D  K  K  V  V  E  Y  T  V  A  K  E  F  A  D  S  L  G  I  P  F  L  E  T

tctgcgaagaacgccagcaacgtcgagcaggctttcctgaccatgccccgccaaatcaaggagcgcatgggcagc  525
 S  A  K  N  A  S  N  V  E  Q  A  F  L  T  M  A  R  Q  I  K  E  R  M  G  S

acgaccgcgaacaacacgaaaccctctgtgcaggttgggcccggacacggcgtcctcgaactcgggcggcggc    600
 T  T  A  N  N  T  K  P  S  V  Q  V  G  P  G  H  G  V  S  S  N  S  G  G  G

tgctgctaa   609
 C  C  *
```